\documentclass[epj]{svjour}
\usepackage{graphics}
\usepackage[dvips]{graphicx}
\let\Otemize =\itemize
\let\Onumerate =\enumerate
\let\Oescription =\description
\def\Nospacing{\itemsep=0pt\topsep=0pt\partopsep=0pt\parskip=0pt\parsep=0pt}
\def\Topspac{\vspace{-0.5\baselineskip}}
\def\Botspac{\vspace{-0.2\baselineskip}}
\newenvironment{Itemize}{\Topspac\Otemize\Nospacing}{\endlist\Botspac}
\newenvironment{Enumerate}{\Topspac\Onumerate\Nospacing}{\endlist\Botspac}

\newcommand{\lsim}{\,{\buildrel < \over {_\sim}}\,}
\newcommand{\gsim}{\,{\buildrel > \over {_\sim}}\,}
\newcommand{\sqrtsNN}{\sqrt{s_{\scriptscriptstyle{{\rm NN}}}}}
\newcommand{\av}[1]{\left\langle #1 \right\rangle}

\newcommand{\gev}{\mathrm{GeV}}
\newcommand{\tev}{\mathrm{TeV}}
\newcommand{\fm}{\mathrm{fm}}

\newcommand{\cm}{\mathrm{cm}}

\newcommand{\mum}{\mathrm{\mu m}}

\newcommand{\PbPb}{\mbox{Pb--Pb}}
\newcommand{\pPb}{\mbox{p--Pb}}

\newcommand{\RAA}{R_{\rm AA}}
\newcommand{\RDh}{R_{{\rm D}/h}}
\newcommand{\pt}{p_{\rm t}}
\renewcommand{\d}{{\rm d}}
\newcommand{\dEdx}{{\rm d}E/{\rm d}x}
\newcommand{\dNdy}{{\rm d}N_{\rm ch}/{\rm d}y}

\newcommand{\ccbar}{\mbox{$\mathrm {c\overline{c}}$}}
\newcommand{\bbbar}{\mbox{$\mathrm {b\overline{b}}$}}

\newcommand{\Dz}{\mbox{$\mathrm {D^0}$}}

\newcommand{\Jpsi} {\mbox{J\kern-0.05em /\kern-0.05em$\psi$}\xspace}

\begin{document}
\title{Open heavy-flavour production in ALICE}

\author{A.~Dainese for the ALICE Collaboration
}                     
%
%
\institute{Dipartimento di Fisica ``G.~Galilei'', Universit\`a degli Studi di Padova, and INFN, Italy}
\date{Received: \today}
%
\abstract{
After a short review of the Physics motivations for the study of 
open heavy flavour production in proton--proton, proton--nucleus 
and nucleus--nucleus collisions at the LHC,
we present results on the expected performance 
of the ALICE experiment for charm and beauty production measurements.
\PACS{25.75.-q, 14.65.Dw, 13.25.Ft} 
} 

\maketitle
\section{Introduction}
\label{intro}

The ALICE experiment~\cite{alicePPR1} will study nucleus--nucleus (AA)
collisions at the LHC, with a centre-of-mass energy $\sqrtsNN=5.5~\tev$ per
nucleon--nucleon (NN) collision for the Pb--Pb system, 
in order to investigate the properties of QCD matter at energy densities of 
up to several hundred times the density of atomic nuclei. Under 
these conditions
a deconfined state of quarks and gluons is expected to be formed.

The measurement of open charm and open beauty production allows to 
investigate the mechanisms of heavy-quark production, propagation and
 ha\-dro\-ni\-za\-tion
in the hot and dense medium formed in high-energy nucleus--nucleus collisions.
The open charm and open 
beauty cross sections are also needed as a reference to measure the effect of
the transition to a deconfined phase on the production of quarkonia.
Heavy-quark production measurements in proton--proton and 
proton--nucleus collisions at the LHC, 
besides providing the necessary baseline for
the study of medium effects in nucleus--nucleus collisions, are interesting 
{\it per se}, as a test
of QCD in a new energy domain.

\section{Heavy-flavour production from pp to AA}

Heavy-quark pairs ($\rm Q\overline Q$) 
are expected to be produced in 
primary partonic scatterings with large virtuality $Q^2>(2m_{\rm Q})^2$ and, 
thus, on small temporal and 
spatial scales, $\Delta t\sim \Delta r\sim 1/Q \lsim 0.1~\fm$
(for $m_{\rm c}=1.2~\gev$).
In nucleus--nucleus reactions, this implies that
the initial production process is not 
affected by the presence of the dense medium formed in the collision.
Given the large virtualities, 
the baseline production cross sections in NN collisions can be 
calculated in the framework of collinear factorization and perturbative QCD
(pQCD)~\cite{hvqmnr}. 
For the estimate of baseline production yields in nuclear collisions
(to be used for performance studies and preparation of the analysis
strategies),
a scaling of the yields with the average number $\av{N_{\rm coll}}$ 
of inelastic NN collisions (binary scaling) is usually assumed:
\begin{equation} 
\d^2 N^{\scriptstyle\rm Q}_{\rm AA(pA)}/\d\pt\d y =
\av{N_{\rm coll}}\times\d^2 N^{\scriptstyle\rm Q}_{\rm pp}/\d\pt\d y\,.
\end{equation}

The expected $\ccbar$ and $\bbbar$ production yields for different collision
systems at the LHC are reported in the first line of 
Table~\ref{tab:xsec}~\cite{notehvq}.
These numbers, assumed as the ALICE baseline, 
are obtained from pQCD calculations at NLO~\cite{hvqmnr}, including 
the nuclear modification of the parton distribution functions 
(PDFs)~\cite{EKS98}
in the Pb nucleus 
(details on the choice of pQCD parameter values and PDF sets can be found 
in~\cite{notehvq}).
Note that the predicted yields have large uncertainties, of about a factor 2,
estimated by varying the values of the calculation parameters.
An illustration of 
the theoretical uncertainty band for the D meson cross section as a function 
of $\pt$ will be shown in section~\ref{exp}, along with the expected 
sensitivity of the ALICE experiment.

\begin{table}[!b]
\caption{Expected $\rm Q\overline Q$ yields per event at the LHC, 
         from NLO pQCD calculations~\cite{notehvq}. 
         For \pPb~and \mbox{Pb--Pb}, 
         PDF nuclear modification 
         is included and $N_{\rm coll}$ scaling is assumed.}
\label{tab:xsec}
\begin{center}
\begin{tabular}{cccc}
\hline
colliding system & pp & p--Pb & Pb--Pb \\
$\sqrtsNN$ & 14 TeV & 8.8 TeV & 5.5 TeV\\
centrality & -- & min. bias & 0--5\% $\sigma^{\rm inel}$\\
\hline
$\ccbar$ pairs & 0.16 & 0.78 & 115 \\
$\bbbar$ pairs & 0.0072 & 0.029 & 4.6  \\
\hline
\end{tabular}
\end{center}
\end{table}

Several effects can determine the breakdown of binary scaling in pA and AA
collisions.
They are usually divided in two classes, that we discuss in the following.

 {\it Initial-state effects}, such as nuclear shadowing, 
      the modification of 
      the parton distribution functions 
      in the nucleus due to gluon recombination 
      at small momentum fraction $x$. 
      Initial-state effects can, at least in principle,
      be studied by comparing proton--proton and proton--nucleus 
      collisions. It has recently been argued that,
      indeed, at LHC energy, gluon recombination may occur even in 
      pp collisions and affect
      the charm production cross section~\cite{dvbek}. 

{\it Final-state effects} in AA collisions, 
      due to the interaction of the produced partons with
      the medium. Partonic energy
      loss in the medium is the main example of such an effect.
Believed to be at the origin of the jet quenching phenomena observed in 
\mbox{Au--Au} collisions at RHIC~\cite{peitzmann}, energy loss is  
     expected to depend
      on the properties of the medium (gluon density and volume) 
      and on the properties of the `probe' parton 
(colour charge and mass).  
Due to the large values of their masses, charm and beauty quarks are 
qualitatively different probes with respect to
 light partons, since, on 
QCD grounds, the in-medium energy loss
of massive partons is expected to be reduced relative to that 
of `massless' partons (light quarks and 
gluons)~\cite{dokshitzerkharzeev,aswmassive,djordjevic}.
In addition to that, since at LHC energy most of the measured 
light-flavour hadrons 
will originate from a gluon parent, heavy-flavour particles, such as 
D mesons, will provide a tool to tag a quark parent.
As pointed out in~\cite{adsw}, the comparison of the high-$\pt$ suppression
for D mesons and for light-flavour hadrons should test the colour-charge 
dependence (quark parent vs. gluon parent) of parton energy loss,
while the comparison for B mesons and for light-flavour hadrons 
should test its mass dependence (massive parent vs. massless parent)
--- in section~\ref{D0} and~\ref{Btoe} we show some predictions 
from~\cite{adsw} and compare them to the expected ALICE sensitivity
for these quenching studies.


\section{Heavy-flavour detection in ALICE}
\label{exp}

The ALICE experimental setup~\cite{alicePPR1,hansake} 
was designed in order to allow the detection
of ${\rm D}$ and ${\rm B}$ mesons in the high-multiplicity environment 
of central \PbPb~collisions at LHC energy, where up to several thousand 
charged particles might be produced per unit of rapidity. 
The heavy-flavour capability of the ALICE detector is provided by:
\begin{Itemize}
\item Tracking system; the Inner Tracking System (ITS), 
the Time Projection Chamber (TPC) and the Transition Radiation Detector (TRD),
embedded in a magnetic field of $0.5$~T, allow track reconstruction in 
the pseudorapidity range $-0.9<\eta<0.9$ 
with a momentum resolution better than
2\% for $\pt<20~\gev/c$ 
and a transverse impact parameter\footnote{The transverse impact parameter,
$d_0$, is defined as the distance of closest approach of the track to the 
interaction vertex, in the plane transverse to the beam direction.} 
resolution better than 
$60~\mum$ for $\pt>1~\gev/c$ 
(the two innermost layers of the ITS are equipped with silicon pixel 
detectors)\footnote{Note that, for pp collisions, the 
impact parameter resolution maybe slightly worse, due to the 
larger transverse size of the beam at the ALICE interaction point.
This is taken into account in the studies presented in the following.}.
\item Particle identification system; charged hadrons are separated via 
$\dEdx$ in the TPC and in the ITS and via time-of-flight measurement in the 
Time Of Flight (TOF) detector; electrons are separated from charged 
pions in the dedicated
Transition Radiation Detector (TRD)~\cite{clemens}, and in the TPC; 
muons are identified in the forward muon 
spectrometer covering in acceptance the range $-4<\eta<-2.5$. 
\end{Itemize}

Detailed analyses~\cite{alicePPR2}, 
based on a full simulation of the detector and of the 
background sources, have shown that ALICE has a good potential to carry out
a rich heavy-flavour Physics programme. 
In section~\ref{D0} we describe the expected performance for the exclusive
reconstruction of ${\rm D^0\to K^-\pi^+}$ 
decays in pp, p--Pb and Pb--Pb collisions, 
and the estimated sensitivity for the study 
of charm energy loss in Pb--Pb collisions. 
In sections~\ref{Btoe} and~\ref{Btomu} we present the perspectives
for the measurement of beauty production in central Pb--Pb collisions 
in the semi-electronic and semi-muonic decay channels. 

For all studies a multiplicity of $\dNdy=6000$
was assumed for central \PbPb~collisions\footnote{This value of the 
multiplicity can be taken as a conservative assumption, since 
extrapolations based on RHIC data predict $\dNdy=2000$--$3000$.}.
We report the results corresponding to the 
expected statistics collected by ALICE per LHC year: 
$10^7$ central (0--5\% $\sigma^{\rm inel}$) \PbPb~events at
$\mathcal{L}_{\rm Pb-Pb}=10^{27}~\cm^{-2}{\rm s}^{-1}$
and $10^9$ pp events at 
$\mathcal{L}_{\rm pp}^{\rm ALICE}=5\times 10^{30}~\cm^{-2}{\rm s}^{-1}$,
in the barrel detectors; the forward muon arm will collect
about 40 times larger samples (i.e.\, $4\times 10^8$ central Pb--Pb events).

\section{Measurement of charm production and in-medium quenching}
\label{D0}

One of the most promising channels for open charm detection is the 
${\rm D^0 \to K^-\pi^+}$ decay (and its charge conjugate) which 
has a branching ratio (BR) of about $3.8\%$.
The expected production yields (${\rm BR}\times\d N/\d y$ at $y=0$) 
for ${\rm D^0}$ (and ${\rm \overline{D^0}}$) 
mesons decaying in a ${\rm K^\mp\pi^\pm}$ pair 
in central 
Pb--Pb (0--$5\%~\sigma^{\rm inel}$) at $\sqrtsNN=5.5~{\rm TeV}$,
in minimum-bias p--Pb collisions at $\sqrtsNN=8.8~{\rm TeV}$ and in pp 
collisions at $\sqrt{s}=14~{\rm TeV}$ are, in the order, 
$5.3\times 10^{-1}$, $3.7\times 10^{-3}$ and $7.5\times 10^{-4}$ per event.

\begin{figure}[!b]
  \begin{center}
  \includegraphics[width=0.5\textwidth]{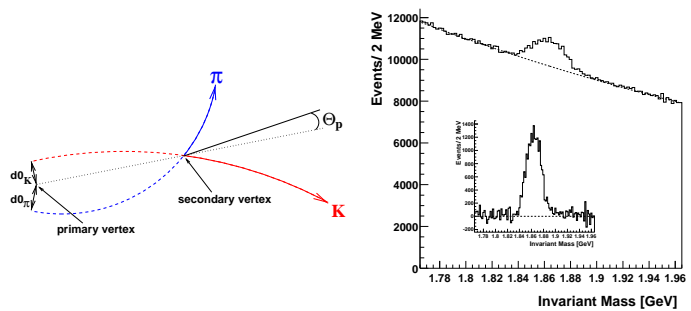}
  \caption{Schematic representation of the ${\rm D^0 \to K^-\pi^+}$ 
           decay (left). 
           ${\rm K\pi}$ 
           invariant-mass distribution corresponding to $10^7$ central Pb--Pb 
           events (right); the background-subtracted distribution is shown 
           in the insert.}
\label{fig:D0combined}
\end{center}
\end{figure}

\begin{figure}[!t]
  \begin{center}
  \includegraphics[width=0.4\textwidth]{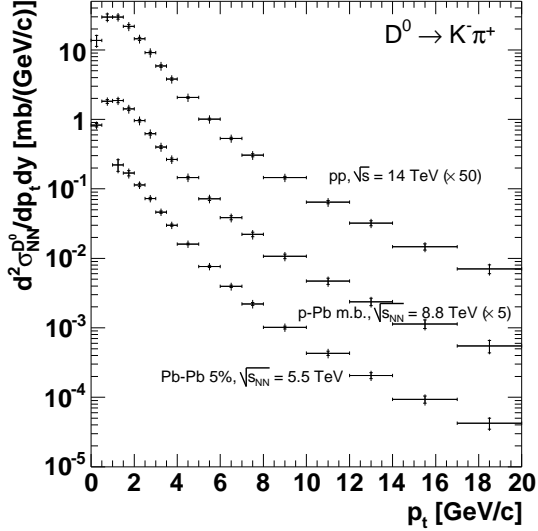}
  \caption{$\pt$-differential cross section per NN collision
             for $\Dz$ production, as expected to be 
             measured with $10^7$ central  
             \PbPb~events
             $10^8$ minimum-bias \pPb~events, 
             and $10^9$ pp minimum-bias events. 
             Statistical (inner bars) and quadratic sum of statistical and 
             $\pt$-dependent 
             systematic errors (outer bars) are shown. A normalization error
             of 9\% for \PbPb, 9\% for \mbox{p--Pb} and 
             5\% for pp is not shown.}
\label{fig:D0pt}
\end{center}
\end{figure}

Figure~\ref{fig:D0combined} 
(left) shows a sketch of the decay: the main feature 
of this topology is the presence of two tracks with impact parameters 
$d_0\sim 100~\mum$. The detection strategy~\cite{D0jpg} to cope with
the large combinatorial background from the underlying event is based on:
\begin{Enumerate}
\item selection of displaced-vertex topologies, i.e. two tracks with 
large impact parameters
and small pointing angle $\Theta_{\rm p}$ 
between the ${\rm D^0}$ momentum and flight-line
(see sketch in Fig.~\ref{fig:D0combined});
\item identification of the K track in the TOF detector;
\item invariant-mass analysis (see $\pt$-integrated
distribution in \PbPb~after selections in Fig.~\ref{fig:D0combined}).
\end{Enumerate}
This strategy was optimized separately for pp, p--Pb 
and \PbPb~collisions, as a 
function of the ${\rm D^0}$ transverse momentum~\cite{thesis,alicePPR2}. 
As shown in Fig.~\ref{fig:D0pt},
the accessible $\pt$ range is $1$--$20~\gev/c$ for \PbPb~and 
$0.5$--$20~\gev/c$ for pp and p--Pb, 
with a statistical error better than 15--20\% 
and a systematic error 
(acceptance and efficiency corrections, 
centrality selection for \PbPb) better than 20\%. More details 
are given in Ref.~\cite{thesis,alicePPR2}.

\begin{figure}[!t]
  \begin{center}
  \includegraphics[width=0.3\textwidth]{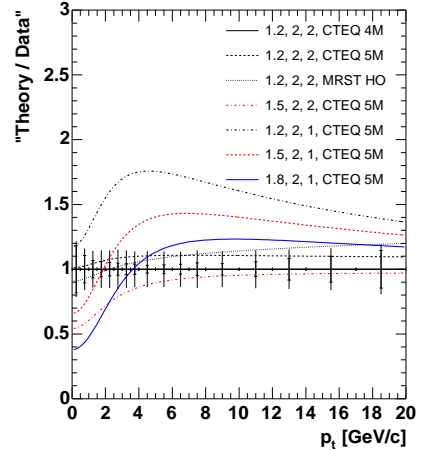}
  \caption{Sensitivity on d$^2\sigma^{\rm D^0}/$d$\pt$d$y$,
           in pp at 14~TeV, compared to 
             pQCD predictions obtained with different sets of input 
             parameters: $m_{\rm c}$ [GeV], 
             the factorization and renormalization scales, in units of 
             $m_{\rm t,c}$,
             and the PDF set. 
             The comparison is shown as a `data/theory' plot.
             Error bars are defined as in Fig.~\ref{fig:D0pt}.}
\label{fig:D0ptcmp}
\end{center}
\end{figure}

For the case of pp collisions, the experimental errors on the 
$\pt$-differential cross section are
expected to be significantly smaller than the current theoretical uncertainty 
band from NLO pQCD calculations (estimated by varying the values of the 
charm quark mass and of the factorization and renormalization scales).
The resulting 'data/theory' plot in 
Fig.~\ref{fig:D0ptcmp} shows that this will allow us to perform 
a sensitive test of the pQCD predictions for charm production at LHC energy.

We studied~\cite{D0epjc} the sensitivity for a comparison of the energy loss 
of charm quarks and of massless partons by considering:
\begin{Itemize} 
\item the {\it nuclear modification factor} of 
D mesons as a function of $\pt$
\begin{equation}
\label{eq:raa}
  R_{\rm AA}^{\rm D}(\pt)\equiv {1 \over \av{N_{\rm coll}}}\times
    \frac{{\rm d}N^{\scriptscriptstyle \rm D}_{\rm AA}/{\rm d}\pt}
       {{\rm d}N^{\scriptscriptstyle \rm D}_{\rm pp}/{\rm d}\pt}\,,
\end{equation}
which is used to characterize the medium-induced high-$\pt$ suppression 
--- in central Au--Au collisions
at RHIC, $\RAA$ is found to be $\simeq 0.2$ for both $\pi^0$ and 
charged hadrons for 
$\pt>4~\gev/c$~\cite{peitzmann};
\item the {\it heavy-to-light 
ratio} of the nuclear modification factors of D mesons and
of charged hadrons:
\begin{equation}
\label{eq:RDh}
   R_{{\rm D}/h}(\pt)\equiv R_{\rm AA}^{\rm D}(\pt)\Big/R_{\rm AA}^h(\pt)\,.
\end{equation}
\end{Itemize}

\begin{figure}[!t]
  \begin{center}
  \includegraphics[width=0.45\textwidth]{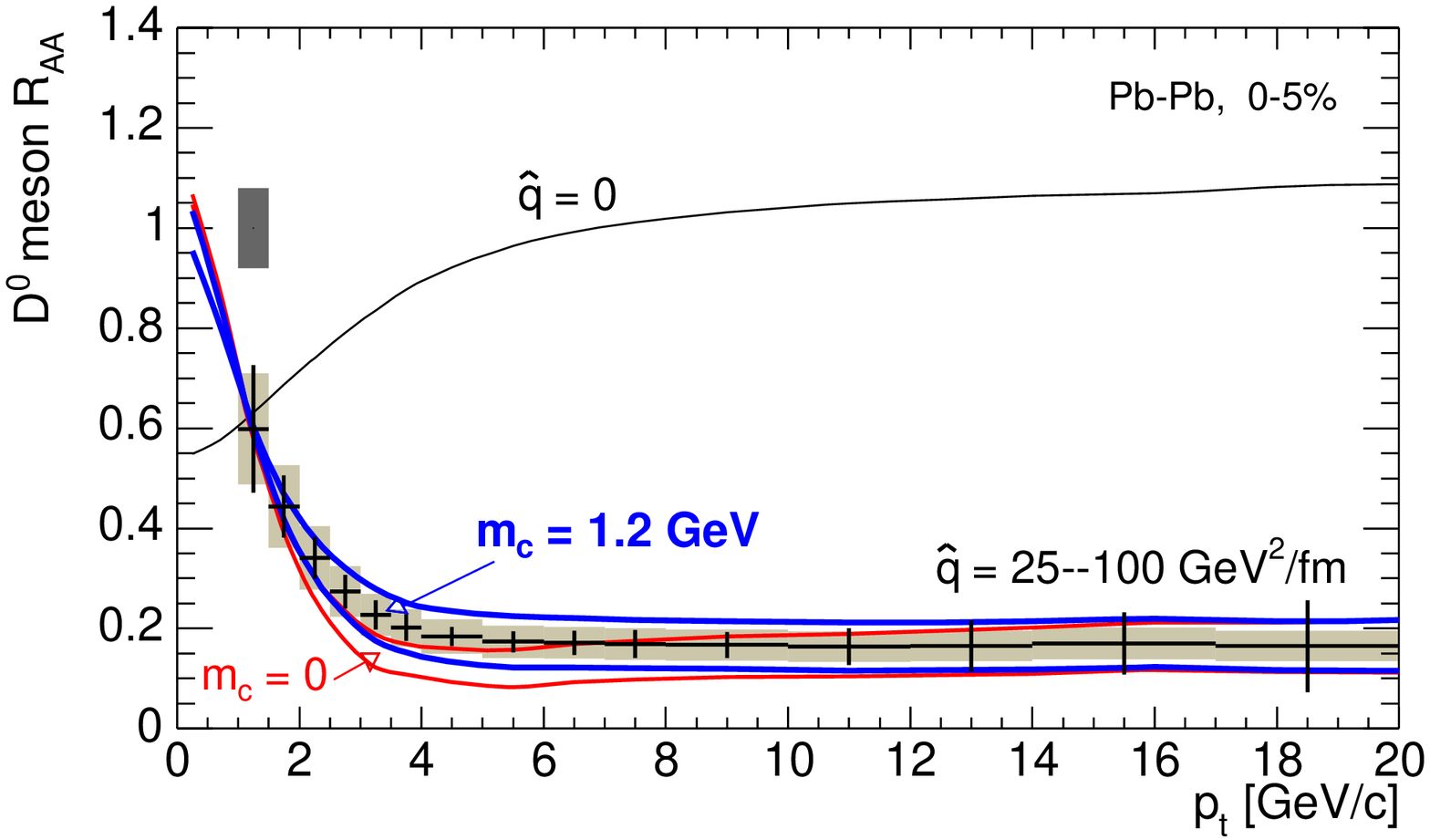}
  \includegraphics[width=0.45\textwidth]{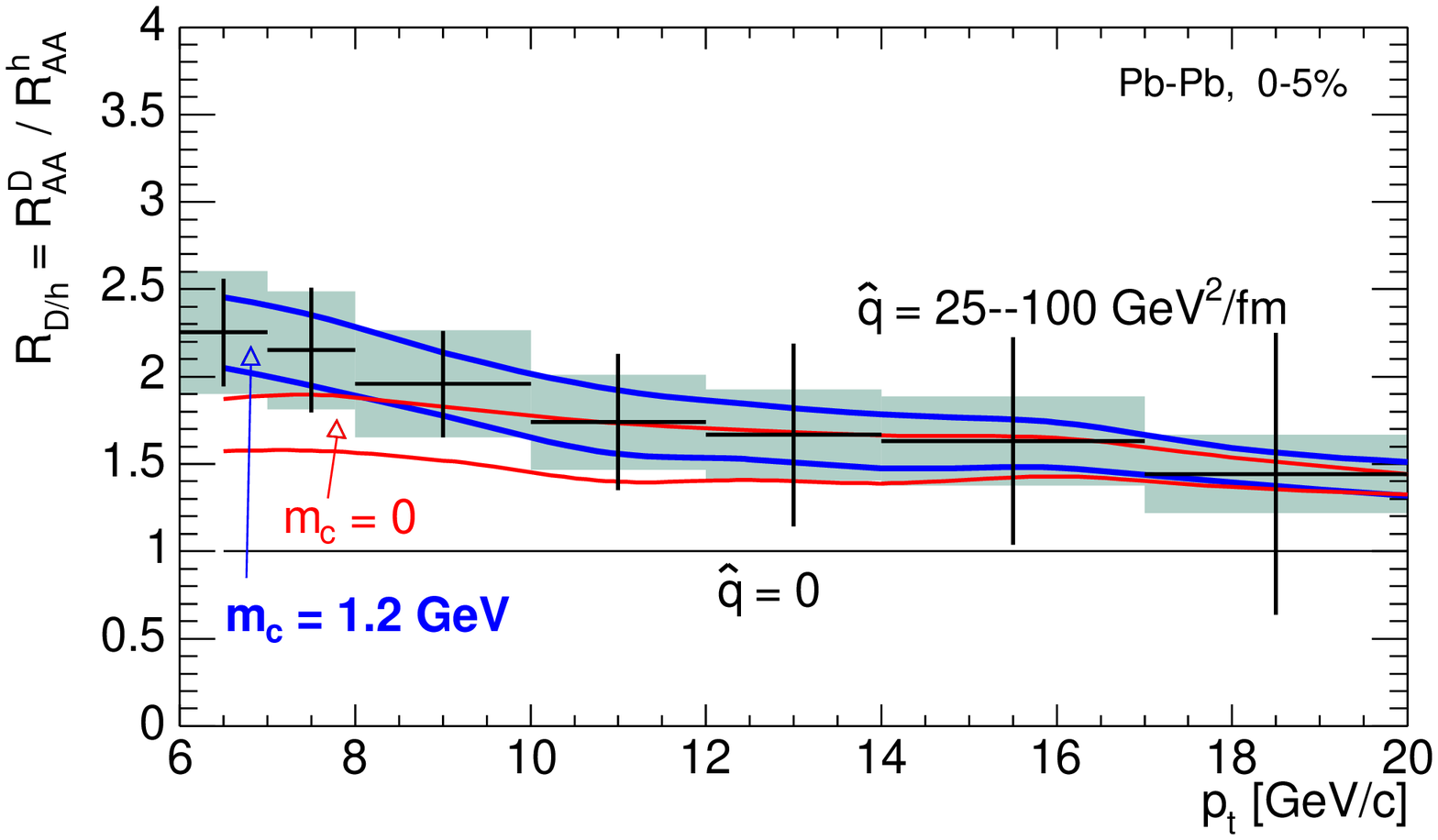}
  \caption{Nuclear modification factor for ${\rm D^0}$ mesons (top) and 
           heavy-to-light ratio of the nuclear modification factors
           for ${\rm D^0}$ mesons and for charged hadrons (bottom). 
           Predictions~\cite{adsw} with and without the effect of the 
           charm mass are shown for the medium density range 
           $\hat{q}=25$--$100~\gev^2/\fm$.
           Errors corresponding to 
           the case `$\hat{q}=50~\gev^2/\fm$ and $m_{\rm c}=1.2~\gev$' 
           are reported: bars = statistical, 
           shaded area = systematic.}
\label{fig:D0quench}
\end{center}
\end{figure}

In Fig.~\ref{fig:D0quench} we compare our estimated sensitivity 
on $R_{\rm AA}^{\rm D}$ and $R_{{\rm D}/h}$
to theoretical calculation results~\cite{adsw} that implement radiative
parton energy loss with medium density described by transport 
coefficient values in the range, $\hat{q}=25$--$100~\gev^2/\fm$,
expected for central Pb--Pb collisions at $\sqrtsNN=5.5~\tev$
on the basis of quenching measurements at RHIC.
The experimental uncertainties,
reported 
in Fig.~\ref{fig:D0quench} 
for the case $\hat{q}=50~\gev^2/\fm$ and $m_{\rm c}=1.2~\gev$, are 
discussed in detail in Refs.~\cite{D0epjc,thesis}. The effect of nuclear
shadowing, introduced via the EKS98 parameterization~\cite{EKS98}, 
is clearly visible 
in the $R_{\rm AA}$ without energy loss for $\pt\lsim 7~\gev/c$. Above this
region, only parton energy loss is expected to affect the nuclear modification 
factor of D mesons. 
The small difference between the theoretical $R_{\rm AA}$ predictions
for $m_{\rm c}=0$ and $1.2~\gev$ indicates that the charm quark 
behaves similarly to a light quark, as far as energy loss is concerned.
Therefore, the enhancement of the heavy-to-light ratio $\RDh$ is a 
sensitive measurement, free of mass effects, to study 
the colour-charge dependence of parton energy. 
As shown by the error bars in the figure, $\RDh$ can be measured with
 good accuracy (as it is a double ratio 
\mbox{(AA/pp)\,/\,(AA/pp)}, some common systematic uncertainties cancel out).

\section{Measurement of beauty production in the semi-electronic 
decay channel}
\label{Btoe}

The production of open beauty can be studied by detecting the 
semi-electronic decays of beauty hadrons, mostly B mesons. 
Such decays have a branching ratio of $\simeq 10\%$ 
(plus 10\% from cascade decays ${\rm b\to c} \to e$, that only populate 
the low-$\pt$ region in the electron spectrum).
The expected yields (${\rm BR}\times\d N/\d y$ at $y=0$)  
for ${\rm b}\to e+X$ plus ${\rm b}\to {\rm c}\,(\to e +X)+X'$ 
in central \PbPb ($0$--$5\%~\sigma^{\rm inel}$) at $\sqrtsNN=5.5~{\rm TeV}$
and in in pp collisions at $\sqrt{s}=14~{\rm TeV}$ 
are $1.8\times 10^{-1}$  and $2.8\times 10^{-4}$ per event, 
respectively.

\begin{figure}[!t]
  \begin{center}
    \includegraphics[width=.4\textwidth]{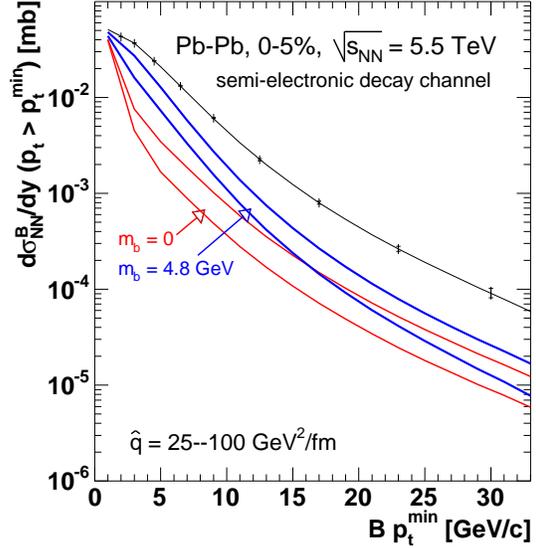}
    \caption{Minimum-$\pt$-differential 
             production cross section per NN collision 
             for B mesons at $y=0$, 
             as expected to be measured from semi-electronic decays with $10^7$
             central Pb--Pb events.
             Statistical errors (inner bars) and quadratic sum of statistical
             and $\pt$-dependent systematic 
             errors (outer bars) are shown.
             A normalization error of 9\% is not shown.
             Suppression 
             predictions~\cite{adsw} with and without the effect of the 
             beauty mass are shown for the medium density range 
             $\hat{q}=25$--$100~\gev^2/\fm$.
             }
    \label{fig:Btoe}
  \end{center}
\end{figure}

The main sources of background electrons are: (a) decays of D mesons; 
(b) neutral pion Dalitz decays $\pi^0\to \gamma e^+e^-$ 
and decays of light mesons (e.g.\,$\rho$ and $\omega$);
(c) conversions of photons in the beam pipe or in the inner detector 
layers and (d) pions misidentified as electrons. 
Given that electrons from beauty have average 
impact parameter $d_0\simeq 500~\mum$
and a hard momentum spectrum, it is possible to 
obtain a high-purity sample with a strategy that relies on:
\begin{Enumerate}
\item electron identification with a combined $\dEdx$ (TPC) and transition
radiation selection, which is expected to reduce the pion contamination 
by a factor $10^4$;
\item impact parameter cut to reject misidentified $\pi^\pm$ and $e^{\pm}$
from Dalitz decays and $\gamma$ conversions 
(the latter have small impact parameter for $\pt\gsim 1~\gev/c$);
\item $\pt$ cut to reject electrons from charm decays. 
\end{Enumerate}
As an example, with $d_0>200~\mum$ and $\pt>2~\gev/c$, the expected statistics
of electrons from b decays is $8\times 10^4$ for $10^7$ central 
\PbPb~events, allowing the measurement of electron-level 
$\pt$-dif\-fe\-ren\-tial 
cross section in the range $2<\pt<18~\gev/c$. 
The residual contamination of about 10\%, accumulated in the low-$\pt$ region, 
of electrons from prompt charm decays, from misidentified charged pions
and $\gamma$-conversion electrons 
can be evaluated and subtracted using a Monte Carlo simulation tuned 
to reproduce the measured cross sections for pions and 
$\rm D^0$ mesons.
A Monte-Carlo-based procedure can then be used to compute,
from the electron-level cross section, the B-level cross section 
$\d\sigma^{\rm B}(\pt>\pt^{\rm min})/\d y$~\cite{alicePPR2}. 
In Fig.~\ref{fig:Btoe} we show this cross section
for central \PbPb~collisions with the estimated 
statistical and systematic uncertainties.
The covered range is $2<\pt^{\rm min}<30~\gev/c$.

The predicted suppression of the B meson $\pt^{\rm min}$-dif\-fe\-ren\-tial 
cross section due to b quark energy loss is also plotted in 
Fig.~\ref{fig:Btoe}. 
The transport coefficient range $25$--$100~\gev^2/\fm$
is considered and the two bands represent the results for 
$m_{\rm b}=0$ and $4.8~\gev$; the two bands are well separated 
up to $\pt^{\rm min}\simeq 15~\gev/c$.
The quenching predictions are shown only for 
illustration, since the study of the B meson suppression 
will have to be performed by using as a reference the cross 
section measured in pp collisions. The sensitivity of this study 
is currently being investigated.

\section{Measurement of beauty production in the semi-muonic decay channel}
\label{Btomu}

Beauty production can be measured also in the ALICE forward muon 
spectrometer, $-4<\eta<-2.5$, analyzing the single-muon $\pt$ distribution
and the opposite-sign di-muons invariant mass 
distribution~\cite{rachid,alicePPR2}.

\begin{figure}[!t]
  \begin{center}
    \includegraphics[width=.45\textwidth]{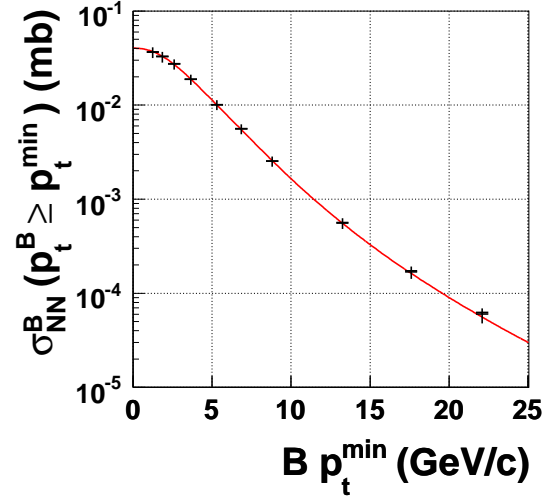}
  \caption{Minimum-$\pt$-differential production 
          cross section per NN collision for B mesons
          with $-4<y<-2.5$ in central Pb--Pb 
           collisions, as expected to be measured 
           from the single-muon data set. 
           Statistical errors (inner bars) corresponding to 
            $4\times 10^8$ events
            and $\pt$-dependent systematic errors (outer bars) 
            are shown. A normalization error of 10\% is not shown.}   
  \label{fig:Btomu}
  \end{center} 
\end{figure}

The main backgrounds to the `beauty muon' signal are $\pi^\pm$, 
$\rm K^\pm$ and charm decays. The cut $\pt>1.5~\gev/c$ is applied to all
reconstructed muons in order to increase the signal-to-background ratio.
For the opposite-sign di-muons, the residual combinatorial background is
subtracted using the technique of event-mixing and the resulting distribution
is subdivided into two samples: the low-mass region, $M_{\mu^+\mu^-}<5~\gev$,
dominated by di-muons originating from a single b quark decay through
$\rm b\to c(\to \mu^+)\mu^-$ ($\rm BD_{\rm same}$), and the high-mass region,  
$5<M_{\mu^+\mu^-}<20~\gev$, dominated by $\bbbar\to\mu^-\mu^+$, with each muon
coming from a different quark in the pair ($\rm BB_{\rm diff}$). 
Both samples have a background 
from $\ccbar\to \mu^+\mu^-$ and a fit is performed to extract the charm- and 
beauty-component yields. The single-muon $\pt$ distribution has three
components with different slopes: K and $\pi$, charm, and beauty decays. 
The first component is subtracted on the basis of the identified hadron spectra
measured in the central barrel. Then, a fit technique allows to 
extract a $\pt$ distribution of muons from beauty decays.
A Monte Carlo procedure, similar to that used for semi-electronic decays, 
allows to extract 
B-level cross sections for the data sets (low-mass $\mu^+\mu^-$, 
high-mass $\mu^+\mu^-$, 
and $\pt$-binned single-muon distribution), 
each set covering a different B-meson $\pt>\pt^{\rm min}$ region. 
The results using only the single muons are 
shown in Fig.~\ref{fig:Btomu}. 
Since only minimal cuts are applied, the reported statistical errors 
(inner bars) are very 
small and the high-$\pt$ reach is excellent.
The main sources of systematic errors (outer bars) are: corrections for 
acceptance and efficiency, subtraction of the background 
muons from charged pion and kaon decays, and fit procedure to 
separate the beauty and charm components.  

\section{Conclusions}

Heavy quarks, abundantly produced at LHC energies, 
will allow to address several physics issues, in pp, pA and AA collisions. 
In particular,
they provide tools to:
probe, via parton energy loss and its predicted colour-charge and mass 
dependences, 
      the dense medium formed in \PbPb~collisions;  
probe, in pp collisions, the pQCD calculations parameters space;
probe the small-$x$ regime of the PDFs, 
      where saturation effects
      are expected to be important.

The excellent tracking, vertexing and particle identification performance 
of ALICE will allow to fully explore this rich phenomenology,
as we have shown with some specific examples on D and B meson 
measurements.

%

\end{document}